%% file: aiptemplate.tex
\documentclass[aip,graphic,preprint]{revtex4-2}
\usepackage{graphicx}
\usepackage{amsmath,mathtools, esint}
\usepackage{xcolor}
\usepackage{amssymb}
\usepackage{soul}
\usepackage{multirow}
\usepackage{array}
\draft 

\begin{document}


\title{Revealing full molecular orientation distributions in organic thin films by nonlinear polarimetry}
\author{Pierre-Luc Thériault}
 \affiliation{Engineering Physics Department, Polytechnique Montréal, Montréal, Québec H3T1J4, Canada}
 \author{Emna Azek}
\affiliation{McGill University, Department of Chemistry, Montréal H3A 0B8, QC, Canada}
\author{Gabriel Juteau}%
 \affiliation{Engineering Physics Department, Polytechnique Montréal, Montréal, Québec H3T1J4, Canada}

 \author{Anagh Mukherjee}
\affiliation{McGill University, Department of Chemistry, Montréal H3A 0B8, QC, Canada}
\author{Heorhii V. Humeniuk}
    \affiliation{McGill University, Department of Chemistry, Montréal H3A 0B8, QC, Canada}
    \altaffiliation{These authors contributed equally}
\author{Zhechang He}
    \affiliation{McGill University, Department of Chemistry, Montréal H3A 0B8, QC, Canada}
\author{Alexandre Malinge}
   \affiliation{Engineering Physics Department, Polytechnique Montréal, Montréal, Québec H3T1J4, Canada}
\author{Dmytro F. Perepichka}

    \affiliation{McGill University, Department of Chemistry, Montréal H3A 0B8, QC, Canada}
\author{Lena Simine}%
\affiliation{McGill University, Department of Chemistry, Montréal H3A 0B8, QC, Canada}
\author{Stéphane Kéna-Cohen}%
 \email{s.kena-cohen@polymtl.ca}
 \affiliation{Engineering Physics Department, Polytechnique Montréal, Montréal, Québec H3T1J4, Canada}

\begin{abstract}
The performance of organic optoelectronic devices is critically dependent on how molecules orient within organic thin films. Yet, standard characterization techniques only reveal the first and second moments of the molecular orientation distribution. This limitation obscures the true molecular arrangement, as diverse distributions can yield identical low-order averages while exhibiting distinct functional properties. Here, we bridge this gap by combining multi-harmonic nonlinear polarimetry (second, third, and fourth harmonic) with the Maximum Entropy Method to reconstruct the probability distribution without any \textit{a priori} assumptions. This allows us to resolve features in the distribution such as asymmetry and bimodality, that remain invisible to conventional probes. Furthermore, we use this method to benchmark molecular dynamics simulations, revealing that these simulations often fail to capture the complex distribution despite correctly predicting the first and second moments. This work transforms molecular orientation from an inferred average into a precise observable, establishing essential validation standards for predictive material design.
\end{abstract}



\date{\today}


\maketitle 

\include{MainText}


%
%

%

\begin{acknowledgments}
 P.L.T. acknowledges support from the Natural Science and Engineering
Research Council of Canada Graduate Scholarship. S.K.C. and L.S. acknowledge support from the Natural Science and Engineering
Research Council of Canada Discovery Grant, the Canada Research Chair Program and the Alliance Quantum Consortium Program (Quantamole). H.V.H. and Z.H acknowledge support from the FRQ Nature et Technologies Postdoctoral Fellowship.

\end{acknowledgments}

\bibliography{aiptemplate}
\newpage
\input{Supplementary}

\end{document}

%% file: MainText.tex
\section*{Statement of significance}
The performance of organic optoelectronic devices, such as OLEDs, depends sensitively on how molecules orient within the various layers. Established characterization techniques can only probe the first two moments of the orientation distribution, obscuring the true complexity of molecular arrangements and limiting the discovery of clear structure-property relationships. Here, we show that nonlinear spectroscopy can be used to reconstruct the full distribution without biased assumptions. This approach reveals hidden structural features, such as bimodality and pronounced asymmetry beyond what established techniques can probe. Crucially, this unprecedented structural detail provides a rigorous benchmark to validate molecular dynamics simulations, accelerating the shift from trial-and-error optimization to the targeted, predictive design of next-generation organic electronic materials.

\section*{Introduction}
\begin{figure}
    \centering
    \includegraphics[width=0.9\linewidth]{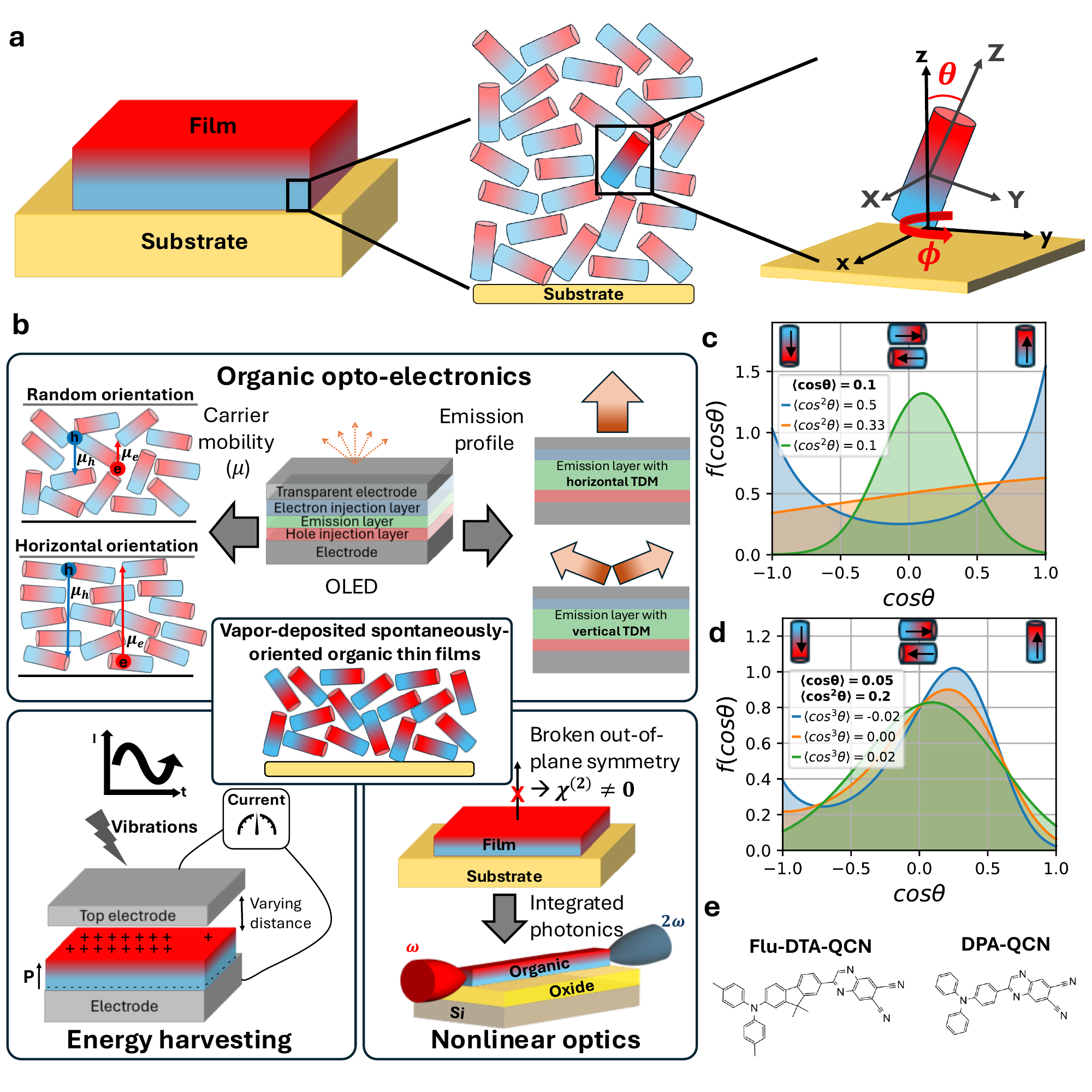}
    \vspace{-20pt}
    \caption{\textbf{Spontaneous molecular orientation andand characterization challenges.}  \textbf{(a)} Schematic of a vapor-deposited film exhibiting spontaneous orientation. The deposition process creates an anisotropic environment, leading to a non-isotropic distribution of molecules in the out-of-plane direction. Due to in-plane isotropy, the orientation of these rod-like molecules can be fully described by a single polar angle, $\theta$. \textbf{(b)} Orientation control is critical across fields: it dictates carrier mobility and OLED outcoupling efficiency in optoelectronics, enables piezoelectricity in energy harvesting\cite{Tanaka_Matsuura_Ishii_2020}, and allows frequency conversion in nonlinear optics by breaking inversion symmetry \cite{Theriault_Malinge_Humeniuk_Bourbonnais-Sureault_Juteau_Martel_Perepichka_Kena-Cohen_2024} ($\chi^{(2)} \neq 0$). \textbf{(c, d)} Ambiguity of low-order moments in describing orientation distributions. \textbf{(c)} Three distinct distributions sharing an identical first moment ($\langle\cos\theta\rangle = 0.1$) but differing in their second moments. \textbf{(d)} Distributions with identical first ($\langle\cos\theta\rangle = 0.05$) and second ($\langle\cos^2\theta\rangle = 0.2$) moments, but distinct third moments ($\langle\cos^3\theta\rangle$). \textbf{(e)} Chemical structures of model compounds Flu-DTA-QCN and DPA-QCN.}
    \label{fig:Fig1}
\end{figure}

The performance of organic electronic devices depends sensitively on how molecules are oriented within each layer\cite{Hofmann_Schmid_Brutting_2021, Pakhomenko_He_Holmes_2023}. In particular, amorphous organic thin films prepared by physical vapor deposition were for a long time assumed to possess a random isotropic molecular arrangement\cite{Yokoyama_Setoguchi_Sakaguchi_Suzuki_Adachi_2010}. However, studies emerging in the early 2000s revealed that during deposition, molecules often adopt non-isotropic orientations\cite{Yokoyama_Setoguchi_Sakaguchi_Suzuki_Adachi_2010, Lin_Lin_Chang_Lin_Wu_Chen_Chen_Chien_Wong_2004, Ito_Washizu_Hayashi_Ishii_Matsuie_Tsuboi_Ouchi_Harima_Yamashita_Seki_2002,Berleb_Brutting_Paasch_2000}, despite being nominally amorphous (\textbf{Fig.~\ref{fig:Fig1}a}). This anisotropy, where the film's properties differ between the in-plane and out-of-plane directions, gives rise to significant macroscopic phenomena\cite{Pakhomenko_He_Holmes_2023, Hofmann_Schmid_Brutting_2021}(\textbf{Fig.~\ref{fig:Fig1}b}). For instance, the collective alignment of polar molecules can generate large surface potentials\cite{Cakaj_Schmid_Hofmann_Brutting_2023, Tanaka_Auffray_Nakanotani_Adachi_2022, Tanaka_2024_boosting}, while the predominantly horizontal orientation of elongated molecules leads to strong optical birefringence and increased out-of-plane mobility\cite{Yokoyama_2011,Yokoyama_Sakaguchi_Suzuki_Adachi_2009, Coehoorn_Lin_Weijtens_Gottardi_vanEersel_2021}.

The profound impact of molecular orientation on device function is now widely recognized\cite{Hofmann_Schmid_Brutting_2021,Pakhomenko_He_Holmes_2023,Ediger_dePablo_Yu_2019,Bagchi_Ediger_2020} (\textbf{Fig.~\ref{fig:Fig1}b}). Anisotropic molecular arrangements have been shown to enhance light outcoupling efficiency \cite{Schmidt_Lampe_Sylvinson_2017, Schmid_Morgenstern_Brutting_2018,Kim2013,Jurow_Mayr_Schmidt_Lampe_Djurovich_Brutting_Thompson_2016}, reduce quenching and improve stability in light-emitting diodes (OLEDs),\cite{Pakhomenko_He_Holmes_2022_polaron, He_Pakhomenko_Holmes_2023,Tanaka_Chan_Nakanotani_Adachi_2024, Rafols-Ribe_Will_Hanisch_Gonzalez-Silveira_Lenk_Rodriguez-Viejo_Reineke_2018} and improve charge transport in organic field-effect transistors (OFETs)\cite{Gundlach_Lin_Jackson_Nelson_Schlom_1997,Sirringhaus_Brown_Friend_Nielsen_Bechgaard_Langeveld-Voss_Spiering_Janssen_Meijer_Herwig_etl._1999, Coropceanu_Cornil_daSilvaFilho_Olivier_Silbey_Bredas_2007}. These functional links have catalyzed the recent emergence of molecular orientation engineering -- the rational design of molecules and processes to achieve specific alignment targets. Current strategies target specific properties, such as maximizing spontaneous orientation \cite{Tanaka_2024_boosting,Tanaka_Auffray_Nakanotani_Adachi_2022,Tanaka_Sugimoto_Nakamura_2025, Wang_Nakano_Hashizume_Hsu_Tajima_2022,Cakaj_Schmid_Hofmann_Brutting_2023} (proportional to the first-order moment of the permanent dipole moment's (PDM) orientation distribution $\langle\cos\theta_{\text{PDM}}\rangle$) or controlling optical anisotropy\cite{Yokoyama_Setoguchi_Sakaguchi_Suzuki_Adachi_2010,Komino_Tanaka_Adachi_2014, Tanaka_Noda_Nakanotani_Adachi_2020, Bishop_Chen_Toney_Bock_Yu_Ediger_2021,Liu_Holmes_2025,Dalal_Walters_Lyubimov_dePablo_Ediger_2015} (proportionnal to the second-order moment of the transition dipole moment (TDM) orientation $\langle\cos^2\theta_{\text{TDM}}\rangle$). However, relying solely on these averages masks the true impact of molecular modifications on the orientation of molecules because a wide variety of distinct orientation distributions can yield identical first and second moments (\textbf{Fig.~\ref{fig:Fig1}c, d}). To decipher the complex intermolecular forces that drive alignment, and thus advance the field from empirical optimization to predictive design, it is essential to move beyond simple averages and reconstruct the underlying orientation distribution.

Characterizing the molecular orientation distribution, however, is a significant challenge \cite{Simpson_Rowlen_2000, Hofmann_Schmid_Brutting_2021,Bruetting_2024, Xu_Jin_Lee_2022}. Because the problem is mathematically underdetermined due to the limited availability of experimental moments\cite{Andersson_Norden_1980} (typically only one or two of the first two moments are available), previous efforts have been forced to rely on simplifying assumptions \cite{Morgenstern_Schmid_Hofmann_Bierling_Jager_Brutting_2018,Yagi_Noguchi_Yokoyama_2024, Simpson_Rowlen_1999,Tronin_Strzalka_Chen_Dutton_Blasie_2000}. Consequently, experimental studies have often defaulted to analytically convenient functions, such as Gaussian or delta functions,  which can obscure the true, potentially complex, nature of the molecular arrangement or even lead to erroneous conclusions. For instance, Simpson and Rowlen have shown that the 'magic interfacial tilt angle' often extracted from second harmonic generation (SHG) measurements is simply an artifact resulting from the assumption of a narrow distribution width \cite{Simpson_Rowlen_1999}.

A recent foray towards solving this challenge has been the use of single-molecule spectroscopy to directly measure the orientation distribution of TDMs molecule-by-molecule\cite{Tenopala-Carmona_Hertel_Hillebrandt_Mischok_Graf_Weitkamp_Meerholz_Gather_2023}. Unfortunately, this technique is restricted to highly dilute guest-host systems and thus cannot probe the crucial intermolecular interactions that govern ordering in most device relevant films \cite{Hofmann_Cakaj_Kolb_Noguchi_Brutting_2025}. Importantly, as a direct probe of the TDM, the technique is insensitive to polar orientation (odd moments of the orientation distribution), which leaves a critical gap in its usefulness for connecting macroscopic properties to their microscopic origins.

Important challenges also exist in the computational domain, hindering validation through powerful predictive tools like molecular dynamics (MD) simulations\cite{acsau_review_comp}. Although MD can provide a microscopic description of film growth during physical vapor deposition\cite{Friederich_Rodin_vonWrochem_Wenzel_2018,Neumann_Danilov_Lennartz_Wenzel_2013,Ishihara_Kaji_2025}, the vast difference in timescales between experiment and simulation often necessitates the use of artificially accelerated parameters, such as deposition rates several orders of magnitude faster than those used in experiment. To bridge this gap, simulation parameters are often calibrated by matching one or two experimentally measured moments. However, given the aforementioned ambiguity, this approach offers weak validation and limited predictive power; multiple simulated realities could be consistent with the same  experimental data. A robust experimental method to determine the orientation distribution with a greater level of details or certainty is therefore critically needed to properly calibrate these models and unlock their potential for predictive materials design.

In this study, we bridge these experimental and computational gaps by introducing a powerful method to reconstruct the molecular orientation distribution with minimal assumptions. We demonstrate that by using nonlinear polarimetry to measure second, third, and fourth-harmonic generation signals, we can extract higher-order moments of the distribution. When combined with a maximum entropy algorithm\cite{Mead_Papanicolaou_1984}, these moments provide sufficient constraints to accurately reconstruct the most probable orientation distribution\cite{Andersson_Norden_1980}. Our approach provides an unprecedented level of detail, establishing a rigorous benchmark for validating theoretical models and transforming molecular orientation from an inferred property into a directly measurable quantity, thereby paving the way for more precise materials design and device optimization.

\section*{Results} 

\subsection*{Framework of the Method}
Our reconstruction technique is simplest when applied to rod-like molecules, for which orientation can be described by a single angle, $\theta$, between the long molecular axis and the substrate normal (\textbf{Fig. \ref{fig:Fig1}a}). For such molecules, the PDM and TDM are collinear with the principal molecular axis ($Z$-axis), and the hyperpolarizability tensors of these molecules possess only one non-zero element, the one where all indices are aligned with the Z-axis\cite{Singer_Kuzyk_Sohn_1987} (\textit{e.g.} $\beta_{ZZZ}$ for the first hyperpolarizability). Given that azimuthal symmetry ($C_{\infty v}$) is inherent to most evaporated and spin-coated films\cite{Yokoyama_Setoguchi_Sakaguchi_Suzuki_Adachi_2010, Pakhomenko_He_Holmes_2023}, the orientation distribution $f$ can be taken to be independent of the azimuthal angle and only a function of the out-of-plane component, $\cos \theta$. Consequently, $f(\cos\theta)$ is referred to as the 'out-of-plane distribution'.

The method relies on the fact that nonlinear optical susceptibilities, $\chi^{(n)}$, are the ensemble average of molecular hyperpolarizabilities over the orientation distribution\cite{Kuzyk_Singer_Stegeman_2013} (SI S1) . For example, for second-harmonic generation in a $C_{\infty v}$ symmetric film with rod-like molecules, the three independent non-zero susceptibility elements are given by\cite{Singer_Kuzyk_Sohn_1987}:

\begin{align}
\label{eq: Susceptibility}
\chi^{(2)}_{zzz}&= (N \beta_{ZZZ})  (l_z^{2\omega} l_z^{\omega} l_z^{\omega}) (\langle cos^3\theta \rangle) \\
\chi^{(2)}_{zxx}&=(N\beta_{ZZZ}) (l_z^{2\omega} l_x^{\omega} l_x^{\omega})(\frac{\langle cos\theta \rangle-\langle cos^3\theta \rangle}{2})\\
\chi^{(2)}_{xzx}&= (N\beta_{ZZZ}) (l_x^{2\omega} l_z^{\omega} l_x^{\omega})(\frac{\langle cos\theta \rangle-\langle cos^3\theta \rangle}{2})
\end{align}
where $N$ is the molecular number density, $l^\omega _U$  is the local field factors at frequency $\omega$, and the uppercase and lowercase indices ($X,Y,Z$ and $x,y,z$) refer respectively to the molecular and laboratory frame (\textbf{Fig. \ref{fig:Fig1}a}). Critically, these susceptibilities are directly proportional to the first and third-order moments of the orientation distribution, $\langle\cos\theta\rangle$ and $\langle\cos^3\theta\rangle$. We note that $\chi^{(2)}_{xzx}$ and $\chi^{(2)}_{zxx}$ differ solely by their local field factor and are therefore not truly independent. 

A variety of techniques have been developped for measuring nonlinear susceptibilities\cite{Kuzyk_Dirk_1998}. If low-order moments (such as $\langle\cos\theta\rangle$) are known from  separate measurements, results from the nonlinear characterization can then be used to extract the remaining unknown moment (such as  $\langle\cos^3\theta\rangle$), and the in-film hyperpolarizability. All other quantities, namely the number density $N$ and the different  $l^\omega _U$ terms, are experimentally accessible or can be calculated (SI S2). \textbf{Table \ref{tab:averages}} lists the orientational averages for the non-zero susceptibility components of orders 2, 3, and 4. Inspection of these expressions reveals a sequential hierarchy: if all moments of the same parity and order less than $(m+1)$ are known, one can systematically extract the $(m+1)^{th}$ moment from a $\chi^{(m)}$ characterization experiment. For instance, knowledge of the first ($\langle\cos\theta\rangle$) and third  ($\langle\cos^3\theta\rangle$) moments enables the determination of the fifth moment $\langle\cos^5\theta\rangle$ from a $\chi^{(4)}$ measurement.

\begin{table}[ht]
\centering
\begin{tabular}{ccc}
\hline
  $\mathbf{\chi^{(m)}}$  &\textbf{Indices} &\textbf{Orientation average}   \\
\hline
\multirow{2}{*}{$\chi^{(2)}$}  & xxz &$\frac{1}{2}(\langle cos\theta \rangle-\langle cos^3\theta \rangle)$\\
 & zzz &$\langle cos^3\theta \rangle$\\\hline
\multirow{3}{*}{$\chi^{(3)}$}  & xxxx &$\frac{3}{8}(1-2\langle cos^2\theta \rangle+\langle cos^4\theta \rangle)$\\
 & xxzz &$\frac{1}{2}(\langle cos^2\theta \rangle-\langle cos^4\theta \rangle)$\\
 &zzzz &$\langle cos^4\theta \rangle$\\\hline
 \multirow{3}{*}{$\chi^{(4)}$}  & xxxxz &$\frac{3}{8}(\langle cos\theta \rangle-2\langle cos^3\theta \rangle+\langle cos^5\theta \rangle)$\\
 & xxzzz &$\frac{1}{2}(\langle cos^3\theta \rangle-\langle cos^5\theta \rangle)$\\
 &zzzzz &$\langle cos^5\theta \rangle$\\\hline
\end{tabular}
 \caption{Orientational average terms for different non-zero terms of the second, third and forth-order susceptibilities. These orientation terms are valid for any permutations of the indices (\textit{e.g.} $\chi^{(3)}_{xxzz}$ has the same orientation term as  $\chi^{(3)}_{zxxz}$) for a $C_{\infty,v}$ symmetry thin film of rod-like molecule. Note that, because of the $C_{\infty,v}$ symmetry, the x direction is equivalent to the y direction.   }
    \label{tab:averages}
\end{table}

Determining a probability distribution from a finite set of its moments is a classic inverse problem\cite{Schmudgen_2017book}. While knowledge of all moments uniquely defines a distribution on a finite interval, any real experiment can only provide a small subset of moments. To select the most physically plausible distribution that agrees with the measured moments, we employ the Maximum Entropy Method \cite{Mead_Papanicolaou_1984,Bandyopadhyay_Bhattacharya_Biswas_Drabold_2005} (MEM). The MEM solves for the distribution that maximizes the information entropy, subject to the constraints imposed by the experimental moments. This approach yields the "flattest" or most random distribution consistent with the data, thus avoiding the introduction of artificial features or assumptions. Moreover, from a numerical optimization perspective, the maximum entropy method is intrinsically regularized, which means that sharp non-physical features are penalized\cite{Mead_Papanicolaou_1984, Sivia_Skilling_2006}. The specific details of our MEM implementation are provided in the Methods section.

\subsection*{Experimental Determination of the Orientation Moments}

\begin{figure}[ht]
    \centering
    \includegraphics[width=0.8\linewidth]{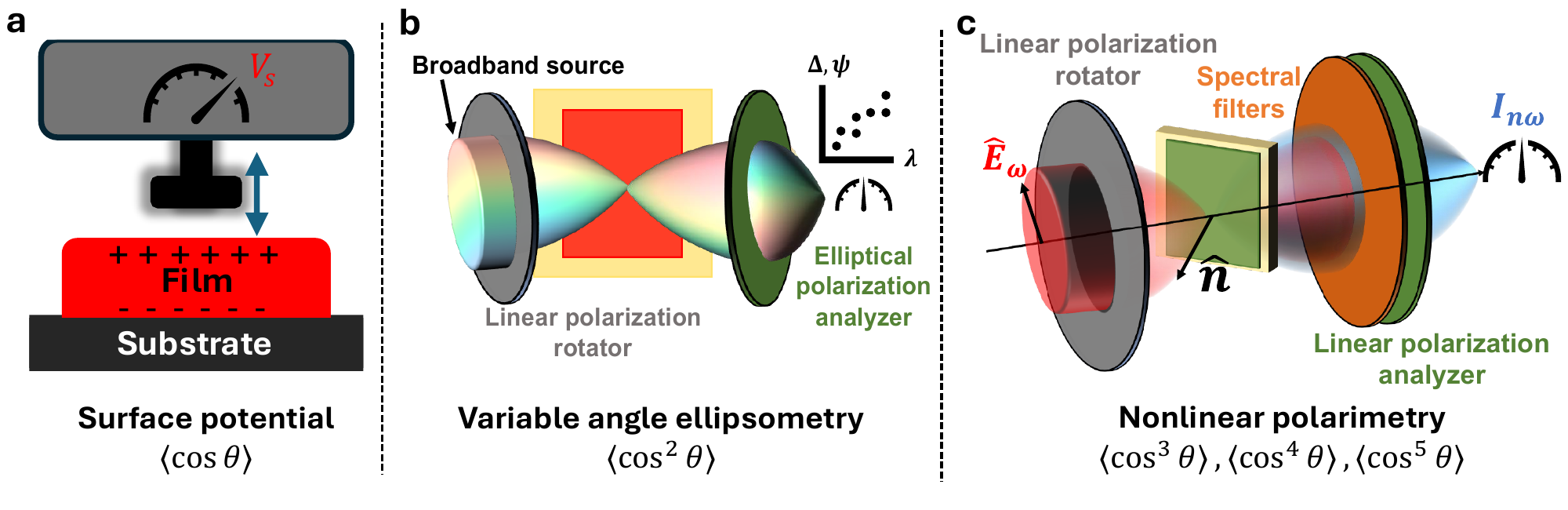}
    \caption{\textbf{ Experimental methods for determining molecular orientation moments.} Three complementary techniques are employed to comprehensively characterize molecular orientation. \textbf{(a)} Surface potential measurements directly probe the net polar order ($\langle\cos\theta\rangle$) of the film. \textbf{(b)} Variable angle spectroscopic ellipsometry measures changes in light polarization (described by ellipsometric parameter $\Delta$ and $\Psi$) to determine the optical anisotropy, from which the second-order moment ($\langle\cos^2\theta\rangle$) is derived. \textbf{(c)} Nonlinear polarimetry measures the intensity of generated harmonic light ($I_{n\omega}$) of two complementary polarization (selected by the analyzer) as a function of the incident polarization (controlled by the polarization rotator) to extract higher-order moments ($\langle\cos^3\theta\rangle$, $\langle\cos^4\theta\rangle$, $\langle\cos^5\theta\rangle$).}
    \label{fig2}
\end{figure}

To illustrate the power of our method, we measure the orientation of two highly polar, rod-like molecules (\textbf{Fig.~\ref{fig:Fig1}e}): 2-(4-(diphenylamino)phenyl)quinoline-6,7-dicarbonitrile (DPA-QCN\cite{footnote}) and 2-(7-(di-p-tolylamino)-9,9-dimethyl-9H-fluoren-2-yl)quinoxaline-6,7-dicarbonitrile (Flu-DTA-QCN), which were previously reported to show spontaneous orientation polarization and strong second-order nonlinearities\cite{theriault2025molecularengineeringenhancedsecondorder}. By performing nonlinear polarimetry up to fourth order and combining the results with standard techniques for the two first moments\cite{Hofmann_Schmid_Brutting_2021} (\textbf{Fig.~\ref{fig2}}), we can experimentally determine the first five moments of the molecular orientation distribution.

The first two moments describe the average polar (or PDM) orientation, $\langle\cos\theta\rangle$, and the average molecular axis (or TDM) alignment, $\langle\cos^2\theta\rangle$. These were determined \cite{Hofmann_Schmid_Brutting_2021}  using surface potential measurements and variable angle spectroscopic ellipsometry(VASE), respectively. Surface potential measurements yielded large thickness-normalized values of $85 \pm 8$\,mV/nm for Flu-DTA-QCN and $90 \pm 9$\,mV/nm for DPA-QCN. These correspond to first moments\cite{theriault2025molecularengineeringenhancedsecondorder} of $0.042 \pm 0.004$ and $0.034 \pm 0.003$, respectively (\textbf{Table~II}), confirming a small but distinct net polar order in both films.

\begin{figure}
    \centering
    \includegraphics[width=0.8\linewidth]{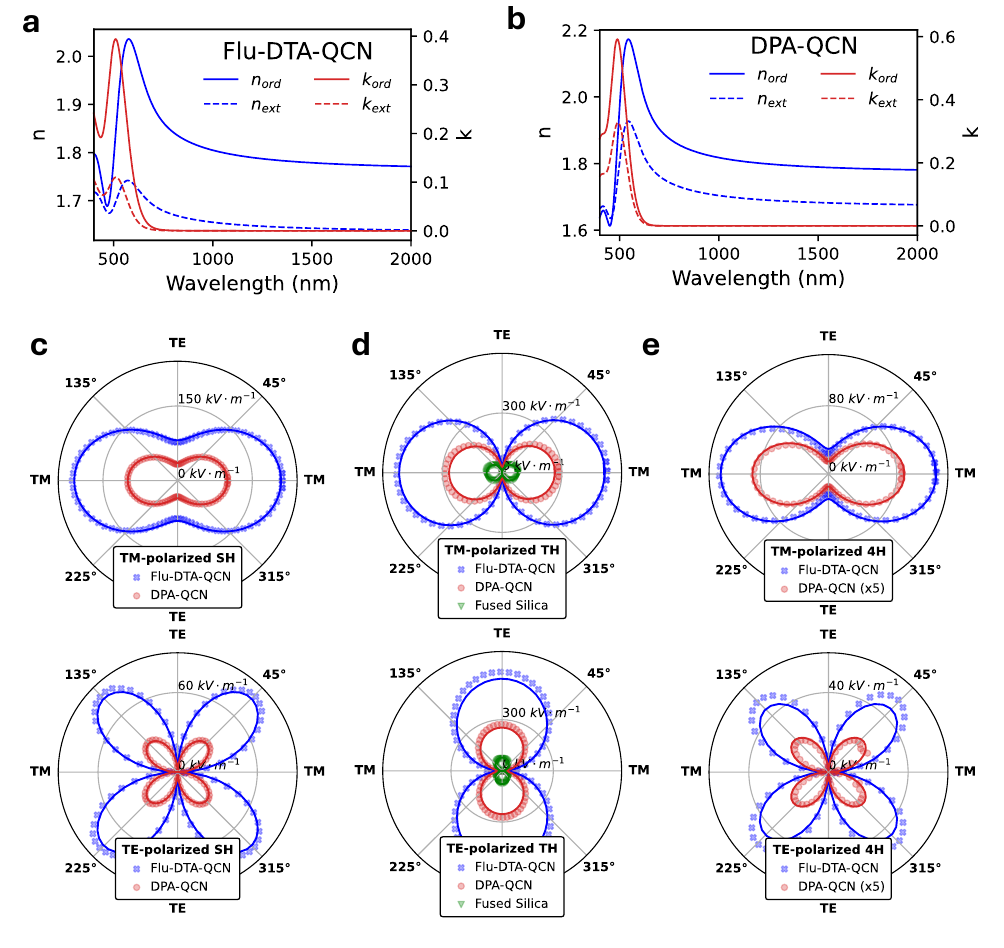}
   \caption{\textbf{Optical and nonlinear characterization of the molecular films.} \textbf{(a, b)} Wavelength-dependent ordinary (\textit{ord}) and extraordinary (\textit{ext}) refractive indices ($n$) and extinction coefficients ($k$) for Flu-DTA-QCN and DPA-QCN films. The significant difference between the in-plane (\textit{ord}) and out-of-plane (\textit{ext}) components confirms a strong uniaxial optical anisotropy in both films. \textbf{(c-e)} Nonlinear polarimetry data for (c) SHG, (d) THG, and (e)FHG . The plots show the generated harmonic electric field magnitude for transverse magnetic (TM, top) and transverse electric (TE, bottom) polarizations. Experimental data (points) are shown alongside the best-fit curves (solid lines) from our nonlinear transfer matrix model. Measurements were performed on films with thicknesses of 77\,nm (Flu-DTA-QCN) and 106\,nm (DPA-QCN). The polarimetry data shown were collected at incidence angles of 50\textdegree, 30\textdegree, and 60\textdegree with on-axis peak intensities of 41, 101, and 210\,GW/cm$^2$ for SHG, THG, and FHG, respectively. For THG (d), the non-negligible contribution from the fused silica substrate (green circles) is shown for comparison and accounted for in the model.}
\label{fig:optical_data}
\end{figure}

The refractive index (\textbf{Fig. \ref{fig:optical_data}a,b}) extracted from the VASE data reveals that the films possess significant birefringence. Indeed, both the real (n) and imaginary (k) components show large differences between their ordinary (in-plane) and extraordinary component (out-of-plane). The second moments (\textbf{Table~II}), derived from the difference in extinction coefficients\cite{Hofmann_Schmid_Brutting_2021}($k_e - k_o$), show that both molecules adopt a predominantly horizontal orientation ($\langle\cos^2\theta\rangle < 1/3$), which is more pronounced for Flu-DTA-QCN. These initial measurements confirm that, while both molecules spontaneously orient during deposition, their underlying distributions are distinct.

\subsection*{Probing Higher-Order Moments with Nonlinear Optics}

To extract higher moments of the distributions, we performed second, third, and fourth-harmonic generation (SHG, THG, FHG) polarimetry on the two materials. Nonlinear polarimetry, shown in \textbf{Fig. \ref{fig2}}, relies on illuminating the sample at an oblique angle with intense, linearly polarized laser pulses at a fundamental frequency ($\omega$) and collecting the transmitted light, containing the generated harmonic signals ($2\omega$, $3\omega$, and $4\omega$). Spectral filters are used to isolate the desired harmonic frequencies from the fundamental pump beam. During the measurement, the polarization of the pump beam is systematically rotated using a half-wave plate to probe different elements of the sample's nonlinear susceptibility tensors. For each input polarization, a calibrated detector measures the intensity of the generated harmonic signal after it passes through a linear polarization analyzer\cite{Hermans_Kieninger_Koskinen_Wickberg_Solano_Dendooven_Kauranen_Clemmen_Wegener_Koos}.

We fit the nonlinear polarimetry data using a nonlinear transfer matrix model that fully accounts for the effects of multiple reflections, anisotropy and absorption losses within the thin film \cite{Bethune_1991}.

By constraining the lower-order moments with values determined from independent measurements (\textit{e.g.}, $\langle\cos\theta\rangle$ and $\langle\cos^2\theta\rangle$), we effectively reduce the problem to two free parameters(see SI S3): the hyperpolarizability (\textit{e.g.}, $\beta_{zzz}$ for SHG) and an unknown moment in the sequence (e.g., $\langle\cos^3\theta\rangle$ for SHG). A global optimization algorithm then identifies the specific values that best reproduce the multi-angle polarimetry dataset (SI S4), allowing for an unambiguous determination of this high-order moment.

The nonlinear polarimetry data and the corresponding fits are shown in  \textbf{Fig.~\ref{fig:optical_data}c-e}. Our nonlinear transfer matrix model shows excellent agreement with the measurements, accurately reproducing the complex polarization-dependent patterns for both molecules. The model explicitly accounts for experimental factors, such as the non-negligible contribution from the fused silica substrate to the third-harmonic signal. From these rigorous fits, we sequentially extracted the third, fourth, and fifth moments of the orientation distribution. The complete set of moments, along with their uncertainties, is compiled in \textbf{Table~\ref{tab:moments}}.

\begin{table}[ht]
    \centering
    \begin{tabular}{ccc}\hline
       \textbf{Moment}  & \textbf{Flu-DTA-QCN} & \textbf{DPA-QCN} \\ \hline 
    $\langle cos \theta \rangle $ &$ 0.042\pm0.004$ & $ 0.034\pm0.003$ \\  $\langle \cos^2 \theta \rangle $ &$ 0.120\pm0.002$& $ 0.217\pm0.004$\\
    $\langle \cos^3 \theta \rangle$ &$ 0.0160\pm0.0003$ & $ 0.0181\pm0.0002$\\ 
     $\langle \cos^4 \theta \rangle$ &$ 0.043\pm0.006$ & $ 0.124\pm0.008$\\
      $\langle \cos^5 \theta \rangle$ &$ 0.0053\pm0.0002$ & $ 0.0089\pm0.0001$\\\hline
    \end{tabular}
    \caption{The first five experimentally determined moments of the molecular orientation distribution for Flu-DTA-QCN and DPA-QCN.}
\label{tab:moments}

\end{table}

\subsection*{Reconstructing the distribution from experimental moments. }

\begin{figure}[ht]
    \centering
    \includegraphics[width=0.8\linewidth]{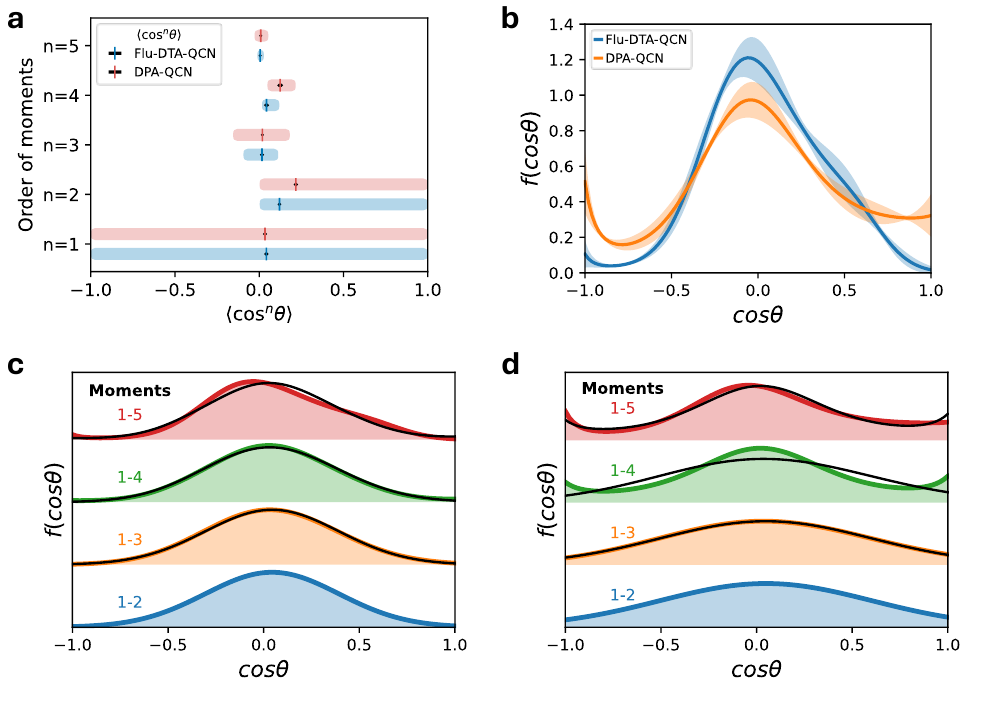}
    \caption{\textbf{Reconstruction and refinement of the molecular orientation distribution.} \textbf{(a)} The first five experimental moments for Flu-DTA-QCN and DPA-QCN. Shaded regions indicate the theoretical bounds for each moment, which are constrained by the values of lower-order moments. \textbf{(b)} Final orientation distributions, $f(\cos\theta)$, obtained using the full set of five moments. The uncertainty (standard deviation), shown as a shaded area, is quantified via a Monte Carlo approach. \textbf{(c, d)} Convergence of the distribution for (c) Flu-DTA-QCN and (d) DPA-QCN. The plots show how the inclusion of additional moments (from the bottom, using moments 1-2, up to the top, using moments 1-5) progressively refines the shape of the reconstructed distribution. The black line in each plot represents the distribution from the preceding set of moments for direct comparison.}
    \label{fig4:Results}
\end{figure}

We can show that the value of any given moment is inherently constrained by the values of the lower-order moments, creating a set of mathematical bounds\cite{Schmudgen_2017} (SI S5). As shown in \textbf{Fig.\ref{fig4:Results}}, all of our experimentally determined moments lie within their theoretical bounds, confirming physical consistency; the tightness of these bounds makes this agreement a stringent internal validation test, substantially strengthening confidence in our multi-technique measurement approach.

From these moments, orientation distributions can be readily extracted with our MEM optimization (see Methods). The extracted distributions for films of both molecules are shown in \textbf{Fig. \ref{fig4:Results}b}. The extracted distributions exhibit key differences. The distribution for DPA-QCN is notably wider, consistent with its larger second moment, and possess significant "tails". Another notable feature relates to the shape of the tails. The rightmost tail, corresponding to a vertical "up" orientation, extends smoothly and monotonically, the leftmost tail decreases  until $\cos\theta=-0.8$ then increases quite sharply. Interpreted in terms of vertical versus horizontal configuration, this non-monotonic behavior indicates a bimodal distribution, with peaks corresponding to molecules oriented horizontally and vertically (see SI S6). In contrast, the distribution for Flu-DTA-QCN is narrower but displays a pronounced asymmetric shoulder on the right side (positive $\cos\theta$). We stress that other than entropy maximization and the experimentally determined moments, no other assumptions about the shape of the distributions are made. The orientation distribution arises solely from entropy maximization and the measured moments.

While the general features of the distributions could be inferred from the first two moments alone, the unique value of our method lies in the refinement provided by higher-order moments, as shown in \textbf{Fig.4c,d}. Interestingly, the inclusion of the third moment, $\langle\cos^3\theta\rangle$, produces only a minor change from the distribution inferred from the first two moments. This is not because the third moment is unimportant, but rather because the maximum entropy distribution predicted from $\langle\cos\theta\rangle$ and $\langle\cos^2\theta\rangle$ coincidentally possesses a third moment very close to our experimentally measured value for these specific molecules. In contrast, the fourth and fifth moments introduce significant, non-trivial features. For DPA-QCN, the inclusion of $\langle\cos^4\theta\rangle$ reveals its subtle bimodal character. For Flu-DTA-QCN, the addition of $\langle\cos^5\theta\rangle$ is necessary to resolve the pronounced shoulder on the right side of the distribution. This stepwise refinement highlights the diagnostic value of higher-order moments: they are required to capture the full complexity of molecular orientation distributions in thin films.

\subsection*{Molecular dynamics simulation of the orientation distribution }

An important application of our technique is its use as a rigorous calibration tool for MD simulations. To date, such models have primarily been used to rationalize observed molecular behaviors \textit{a posteriori} \cite{Friederich_Rodin_vonWrochem_Wenzel_2018,Neumann_Danilov_Lennartz_Wenzel_2013,Ishihara_Kaji_2025}. While simulations can reproduce experimental trends, their reliance on tunable parameters (\textit{e.g.}, force field selection, deposition rate, temperature) can lead to superficial agreement with low-order moments that fails to reflect the true underlying physics or full orientation distribution.

\begin{figure}[ht]
    \centering
    \includegraphics[width=0.4\linewidth]{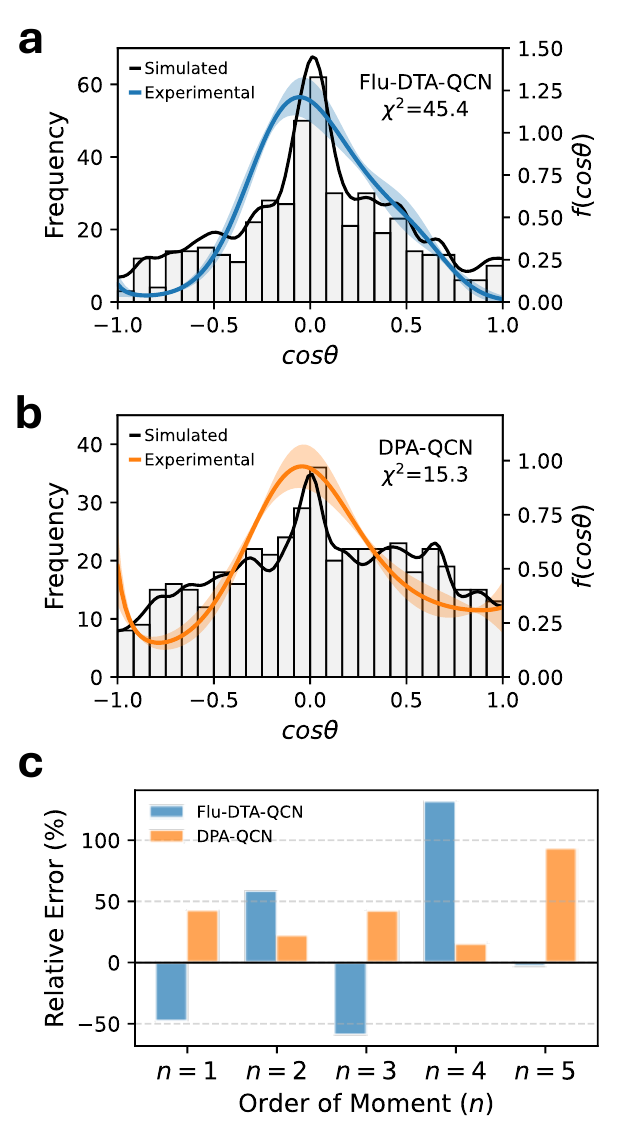}
\caption{\textbf{Comparison between experimentally reconstructed and molecular dynamics simulated orientation distributions.} \textbf{(a, b)} Overlay of the experimentally retrieved orientation distributions (colored solid lines) obtained via the Maximum Entropy Method and the distributions obtained from molecular dynamics simulations (black histograms) for \textbf{(a)} Flu-DTA-QCN and \textbf{(b)} DPA-QCN. The shaded regions around the experimental curves represent the uncertainty derived from the Monte Carlo analysis. The $\chi^2$ values indicate the goodness of fit between the simulation and the experiment. The solid black line shows a Gaussian smoothed distribution obtained from the raw simulated orientation data. \textbf{(c)} Relative error between the simulated and experimental moments ($\langle\cos^n\theta\rangle$) for orders $n=1$ to $5$.}
\label{fig:Fig5}
\end{figure}

To illustrate this, we performed MD simulations of the deposition process for our two materials (see Methods and SI S7). \textbf{Fig.~\ref{fig:Fig5}a,b} overlays the experimentally retrieved distributions with the smoothed distributions obtained from MD. The simulations reproduce salient features, such as the narrower distribution of Flu-DTA-QCN compared to DPA-QCN and the presence of an asymmetric shoulder for Flu-DTA-QCN. However, subtle experimental features, such as the population increase near $\cos\theta=-1$ for DPA-QCN, are absent in the simulation, while spurious features appear, such as a shoulder at $\cos\theta=0.5$ for DPA-QCN. Notably, both simulated distributions exhibit a sharp, prominent peak near $\cos\theta=0$ which differs significantly from the experimentally extracted distributions. The goodness-of-fit, quantified by $\chi^2$ values of 45.4 (Flu-DTA-QCN) and 15.3 (DPA-QCN), indicates a quantitatively better agreement for DPA-QCN, despite these discrepancies.

A  quantitative comparison of the moments (\textbf{Fig.~\ref{fig:Fig5}c}) from the simulated distributions (SI S8) with the experimentally obtained moments further reveals the specific limitations of the simulations. We note that the statistical fluctuations inherent to the finite system size of MD simulations do not significantly affect the calculated moments, because the integration process averages out such high-frequency noise. The first two moments ($n=1,2$) agree reasonably well (relative error generally $<60\%$), illustrating that the simulations correctly capture the polarity sign ($\langle\cos\theta\rangle>0$ ) and the general tendency toward horizontal alignment ($\langle\cos^2\theta\rangle<1/3$). However, the final distribution shape is determined by the cumulative contribution of all moments. Consequently, the compounding effect of deviations—which can approach or exceed 100\% for high-order moments—leads to distributions that differ significantly from the experiments. This suggests that while the model captures the dominant alignment, it fails to account for the specific intermolecular interactions that dictate the complex orientation distributions of the films.

\section*{Discussion}

Our results highlight the risks of describing molecular orientation solely using low-order moments. Based on their first moments, DPA-QCN and Flu-DTA-QCN appear nearly indistinguishable in their polarity. A reliance on this single scalar value, as is common in studies focusing on spontaneous orientation polarization, would erroneously suggest that the driving forces for alignment are effectively identical in both systems. Adding the second moment provides the first distinction, revealing that the more compact DPA-QCN is, on average, more vertical, a behavior expected for compact molecular geometries\cite{Yokoyama_2011}. Crucially, however, even these two moments are insufficient to predict the bimodality of the DPA-QCN distribution or the pronounced asymmetry in Flu-DTA-QCN. While elucidating the precise microscopic origin of these features is beyond the scope of this study, the ability to resolve such differences demonstrates that the full distribution contains critical structural information that simple averages inherently discard.

Efforts to resolve orientation distributions trace back to early surface second-harmonic generation studies on monolayers\cite{heinz1982spectroscopy,heinz1983determination}. However, as noted by Simpson and Rowlen, these early works relied heavily on the assumption of narrow distributions, often leading to erroneous conclusions \cite{Simpson_Rowlen_1999, Simpson_Rowlen_2000}. Subsequent efforts in on monolayers and poled polymers attempted to mitigate this by combining linear spectroscopy and SHG to simultaneously access the second and third moments\cite{Park_Kinoshita_Sakai_Yoo_Hoshi_Ishikawa_Takezoe_1998, Simpson_Westerbuhr_Rowlen_2000,Mortazavi_Knoesen_Kowel_Higgins_Dienes_1989,Palazzesi_Stella_DeMatteis_Casalboni_2010}. However, without independent access to the first moment, the orientation distribution remained mathematically unconstrained \cite{Andersson_Norden_1980}. For guest-host polymer films, combinations of linear spectroscopy with Raman spectroscopy have successfully accessed the second and fourth moments\cite{Xu_Jin_Lee_2022,Runge_Saavedra_Mendes_2006,Richard-Lacroix_Pellerin_2013}. However, the inability of this approach to resolve odd moments makes it blind to polar ordering. Even more recent methods, such as combining surface potential measurements with solid-state NMR\cite{Bruetting_2024}, are limited by the need for specific nuclear isotopes and the inaccessibility of high-order odd moments. Among these techniques, our nonlinear optical approach stands out as the most comprehensive, offering simultaneous access to high-order moments of both parities (odd and even), which are essential for an accurate description of the orientation distribution\cite{Andersson_Norden_1980}.

In addition, the ability to measure orientation distributions establishes a rigorous new benchmark for the computational prediction of organic thin film morphologies. As demonstrated by our comparison with MD simulations, reliance on low-order moments alone can mask significant deviations in the underlying molecular arrangement. While the simulations successfully captured the key features of the films, matching the first and second moments with reasonable accuracy, they failed to reproduce the fine structural details and introduced spurious features. Historically, MD force fields have been validated by matching a single experimental scalar (\textit{i.e.}, $\langle\cos\theta\rangle$ and/or $\langle\cos^2\theta\rangle$), a weak constraint that our data show can be satisfied even by incorrect physical models. By validating simulations against a larger set of moments, we provide the constraints necessary to pinpoint specific deficiencies in the interaction potentials or deposition protocols. This establishes a robust feedback loop: experiment validates and refines the MD model, which can then be trusted to predict the properties of novel molecular candidates \textit{in silico}, accelerating the transition from empirical optimization to predictive design.

While we extract moments up to the fifth order, the reconstructed distribution remains an estimate. Technically, accessing even higher-order moments ($\chi^{(n>4)}$) is possible within our framework but practically challenging because the harmonic signal strength diminishes rapidly with nonlinear order\cite{boyd2020nonlinear}. However, the physical necessity of these higher moments is debatable. In harmonic analysis, higher-order moments describe increasingly sharp features\cite{Andersson_Norden_1980}. In amorphous, vacuum-deposited films, such sharp features are physically unlikely\cite{Ediger_dePablo_Yu_2019, Zallen_2007}. The fact that our reconstructed distributions change only subtly upon the inclusion of the highest moments validates this view, indicating that MEM provides a reasonable and robust description of the system even before the full set of moments is available. Unlike model-dependent approaches that force data into an artificial mold (e.g., Gaussian), MEM is inherently flexible. Moreover, the underlying principle of entropy maximization naturally reflects the physics of disordered assemblies. By producing the smoothest distribution consistent with the measured data, MEM avoids the non-physical sharp artifacts associated with truncated high-order expansions\cite{Andersson_Norden_1980}. Consequently, complex features like the asymmetry in Flu-DTA-QCN can be confidently attributed to the data itself rather than a priori assumptions.

Although this study focused on physical vapor-deposited films as a model system, the methodology presented here is readily applicable to any film exhibiting in-plane rotational symmetry, such as those prepared by spin-coating. Nonetheless, despite its comprehensiveness, our framework operates within specific constraints. The current model assumes rod-like molecules where the hyperpolarizability is dominated by a single axial component. While this approximation covers a vast class of conjugated molecules, the underlying formalism can be generalized to accommodate chromophores with significant off-diagonal tensor elements.

Furthermore, the implementation of multi-harmonic polarimetry requires high-power laser sources, making it less suitable for high-throughput screening than linear optical methods. However, the primary value of this technique lies not in routine characterization, but in fundamental discovery. By performing systematic studies on carefully selected model systems, this method can generate the high-fidelity experimental data needed to uncover key intermolecular interactions influencing orientation and calibrate predictive computational tools.

\section*{Conclusion}
In summary, we have developed and demonstrated a robust methodology to reconstruct the molecular orientation distribution within organic thin films, effectively overcoming the ill-posed nature of standard characterization methods. By combining surface potential measurements, spectroscopic ellipsometry, and multi-harmonic nonlinear polarimetry, we determined the first five moments of the orientation distribution for two model polar molecules. This allows for a maximum entropy reconstruction of the probability function $f(cos\theta)$ without relying on idealized shape assumptions, revealing complex structural features such as bimodality and asymmetry, that would be absent from with other recovery tools relying solely on low-order moments. Furthermore, by providing a rigorous experimental "ground truth",  our method serves as a critical validation tool for molecular dynamics simulations, a key step toward enhancing their predictive capability. As the fields exploiting organic thin films continue to mature and expand, the ability to measure and engineer the full orientation landscape will be essential. We anticipate that this distribution-centric approach will become a standard paradigm for elucidating structure-property relationships, ultimately enabling the rational design of next-generation materials with tailored optical, electronic, and polar properties.

\section*{Methods }
\subsection{Materials and film fabrication}
The molecules were synthesized and purified following the protocols detailed in Ref. \citenum{theriault2025molecularengineeringenhancedsecondorder}. Prior to deposition, the bulk powder was ground and degassed in a vacuum chamber for 20 minutes at a power slightly below the sublimation threshold. Thin films were fabricated by thermal evaporation onto fused silica substrates and thermal oxide (100 nm) on Si using an Angstrom Engineering EvoVac deposition chamber (base pressure $< 10^{-6}$ Torr). The deposition rate was monitored using a quartz crystal microbalance and maintained at 0.75 \AA/s for DPA-QCN and 1.5 \AA/s for Flu-DTA-QCN. For the samples discussed in this work, the DPA-QCN film had a thickness of 106 nm and the Flu-DTA-QCN film had a thickness of 77 nm, as determined by spectroscopic ellipsometry.

\subsection{Surface potential measurements}
The spontaneous orientation polarization was characterized by measuring the surface potential of the films using a Trek 320C electrostatic voltmeter. The Giant Surface Potential (GSP) slope, representing the potential build-up per unit thickness (mV nm$^{-1}$), was determined by measuring the potential difference between the deposited film and the bare substrate, normalized by the thickness of the film. Additional details on the experimental procedure and the extraction of the average molecular orientation angle $\langle \cos\theta \rangle$ from the GSP slope can be found in Ref. \citenum{theriault2025molecularengineeringenhancedsecondorder}. The reported uncertainty for the first moment reflects sample-to-sample variability (the standard deviation across three films). This value does not account for uncertainties in the magnitude of the permanent dipole moment and the static permittivity used to compute the moment from surface potential measurements \cite{theriault2025molecularengineeringenhancedsecondorder}.

\subsection{Linear optical properties}
The linear optical constants (refractive index $n$ and extinction coefficient $k$) were determined by variable angle spectroscopic ellipsometry (VASE) combined with oblique-incidence reflection spectroscopy. These measurements were performed on films deposited on a silicon substrate with a 100 nm thermal oxide layer to facilitate the reflection measurements and suppress coherent backside reflections. VASE measurements were performed using a J.A. Woollam RC2 ellipsometer, while reflection spectra were acquired using an Agilent Cary 5000 spectrophotometer.

The datasets were analyzed simultaneously using a multi-sample analysis in the CompleteEase software. The optical constants were modeled using an anisotropic uniaxial approximation, where the ordinary ($n_{ord}$, $k_{ord}$, in-plane) and extraordinary ($n_{ext}$, $k_{ext}$, out-of-plane) components were fitted using a generalized oscillator model. This procedure allowed for the precise determination of film thickness and the quantification of the uniaxial complex refractive index, as detailed in Ref. \citenum{theriault2025molecularengineeringenhancedsecondorder}. The reported uncertainty for the second moment is derived from the confidence intervals given by the VASE fitting procedures.

\subsection{Nonlinear polarimetry}
The nonlinear optical response was characterized via nonlinear polarimetry in a transmission geometry. Incident light pulses ($\tau_{\text{FWHM}} \approx 170$ fs) were generated using an optical parametric amplifier (Light Conversion Orpheus) pumped by a femtosecond laser system (Light Conversion Pharos, $\lambda = 1030$ nm, $\tau_{\text{FWHM}} \approx 260$ fs). 

SHG measurements were performed at a fundamental wavelength of 1550 nm, while THG and FHG measurements used a fundamental wavelength of 1900 nm. For the SHG experiments, the films were probed at incidence angles of 45$^\circ$, 50$^\circ$, 55$^\circ$, and 60$^\circ$, and the generated signal was isolated using a combination of longpass and shortpass filters. For the THG and FHG experiments, the nonlinear signal was spectrally separated from the fundamental beam using a dispersive prism and spatially filtered to suppress the pump and other harmonics. THG measurements were acquired at incidence angles of 0$^\circ$, 30$^\circ$, 45$^\circ$, and 60$^\circ$, while FHG measurements were conducted at 45$^\circ$ and 60$^\circ$. 

The signal intensity was detected using a calibrated CMOS camera. The orientational moments were extracted using a global fit approach that simultaneously modeled the data across all incidence angles for both TE- and TM-polarized nonlinear outputs. The reported uncertainties for the higher-order moments are obtained directly from the confidence intervals of the global fit parameters.

\subsection{Maximum entropy algorithm}
The orientation distribution function $f(\cos\theta)$ was determined numerically by discretizing the variable $x = \cos\theta$ over a uniform grid of $N$ points, $\{x_i\}$, on the interval $[-1, 1]$. The discrete probability distribution, $\{f_i\}$, was found by maximizing the Shannon-Gibbs entropy functional, $S = -\sum_{i=1}^{N} f_i \ln(f_i) \Delta x$, where $\Delta x$ is the grid spacing. This maximization was performed subject to a set of hard equality constraints: the normalization of the distribution ($\sum_{i=1}^{N} f_i \Delta x = 1$) and the exact reproduction of the $K$ experimentally measured moments ($\sum_{i=1}^{N} x_i^n f_i \Delta x = \langle\cos^n\theta\rangle_{\text{exp}}$ for each moment  $n$). A non-negativity constraint ($f_i \ge 0$) was also enforced. This constrained non-linear optimization problem was solved using the Sequential Least Squares Programming (SLSQP) algorithm.

To quantify the uncertainty in the final distribution arising from experimental errors in the moments, a Monte Carlo simulation was performed. For each of the $M$ iterations, a new set of target moments was generated by sampling from independent normal distributions defined by the experimental mean and standard deviation of each moment. The Maximum Entropy optimization was then solved for this new set of constraints, generating an ensemble of plausible distributions. The final reported distribution is the arithmetic mean of this ensemble, and its uncertainty is given by the standard deviation at each point of the discretized grid.

\subsection{Molecular dynamics simulations}

The thin-film deposition process was simulated using classical molecular dynamics (MD) within the LAMMPS framework \cite{LAMMPS} in an elongated simulation box (10 nm by 10 nm by 20 nm) with two-dimensional periodic boundary conditions in the x and y directions with and a confined z-dimension extending 20 nm to accommodate the film growth. The simulation was initialized with a substrate (10 nm by 10 nm graphene double layer) placed in the x-y plane at the bottom of the box. The molecular dynamics were simulated using modified OPLS force field \cite{ff} (see SI S6 for details), NVT Langevin dynamics were integrated with 1 fs time-step, a damping factor 100 fs, and thermostat temperature set to $0.8T_g$ where $T_g$ is the the material-specific glass transition temperature. For the computational identification of $T_g$ for DPA-QCN and Flu-DTA-QCN films see Section S6. The organic molecules  were initially optimized at the B3LYP/6-31G(d) level and represented via molecular data files generated with molTemplate\cite{moltemplate}, they were introduced into the simulation box from the top with the initial z-component of velocity set to 5 Å/ps towards the substrate, emulating gravitational settling, with new molecules injected every 2 ps. The bottom layer of graphene was frozen to maintain structural stability. Deposition was stopped once the film thicknesses approached 10nm. Post-deposition, the system was equilibrated for 2 ns. The final configurations were used in the analysis of morphological properties with molecules within 4nm from surface removed from consideration. 
The smoothed distributions in \textbf{Fig.~\ref{fig:Fig5}} were obtained from histograms (bin size 0.02) by applying a Gaussian filter with a kernel width of 0.05.

%% file: Supplementary.tex
\newpage
\section{Supplementary Materials}

\subsection*{S1. Connecting Nonlinear Susceptibilities to Orientational Averages}

The macroscopic nonlinear susceptibility of order $n$, $\chi^{(n)}$, is related to the corresponding microscopic hyperpolarizability, $\beta^{(n-1)}$, through a statistical average over the molecular orientation distribution\cite{Kuzyk_Singer_Stegeman_2013}. The general relationship is given by\cite{andrews2002optical,prasad1991introduction}:

\begin{equation}
\label{eq:SusceptibilityGeneral}
\chi^{(n)}_{i_0 i_1 \dots i_n} = N l_{i_0}^{n\omega}\left( \prod_{j=1}^{n} l_{i_j}^\omega \right) \sum_{I_0, \dots, I_n} \langle R_{i_0}^{I_0} R_{i_1}^{I_1} \dots R_{i_n}^{I_n} \rangle \beta^{(n-1)}_{I_0 I_1 \dots I_n}
\end{equation}

\begin{figure}[ht!]
    \centering
    \includegraphics[width=0.3\linewidth]{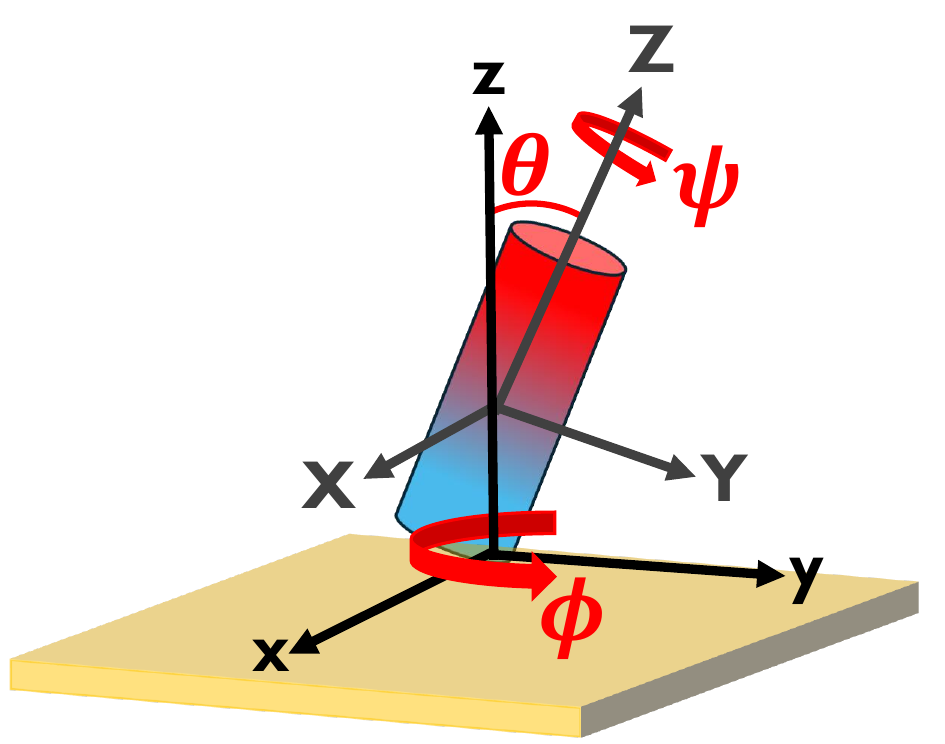}
    \caption{\textbf{Definition of the coordinate system.} Orientation of the molecular reference frame ($X,Y,Z$) with respect to the substrate laboratory frame ($x,y,z$). The transformation is parameterized by a sequence of 3 rotaitons defined with Euler angles: the azimuthal angle $\phi$ (rotation around the substrate normal $z$), the polar tilt angle $\theta$ (rotation around the update X-axis), and the twist angle $\psi$ (rotation around the updated Z-axis).  The order of rotation depicted is consistent with the standard $z$-$X'$-$Z''$ conventions used in descriptions of molecular orientation.}
    \label{fig:SuppRefFrame}
\end{figure}

In this governing equation, the uppercase indices ($I_j = X,Y,Z$) refer to the molecular frame, while the lowercase indices ($i_j = x,y,z$) refer to the laboratory (substrate) frame. The transformation between these frames is accomplished using an Euler rotation matrix ($R_{xyz}^{XYZ}$ given below\cite{Kuzyk_Singer_Stegeman_2013}) with elements $R^U_v$. The brackets, $\langle\dots\rangle$, denote an ensemble average over the entire molecular distribution, which is a function of the Euler angles ($\phi, \theta, \psi$). The angles are described as follow: If the two frames are initially overlapped, a rotation around z correspond to the angle $\phi$, then  a rotation around the updated X coordinate defines $\theta$, then rotation around the
updated Z coordinate defines the angle $\psi$. These angles are shown on Figure \ref{fig:SuppRefFrame}. The equation is also scaled by the molecular number density, $N$, and includes the necessary local field correction factors, $l_{i_J}$.

\begin{align}
    R_{xyz}^{XYZ}=\left(\begin{array}{ccc}
\cos \phi \cos \psi-\sin \phi \cos \theta \sin \psi & -\cos \phi \sin \psi-\sin \phi \cos \theta \cos \psi & \sin \phi \sin \theta  \\
\sin \phi \cos \psi+\cos \phi \cos \theta \sin \psi & -\sin \phi \sin \psi+\cos \phi \cos \theta \cos \psi & -\cos \phi \sin \theta \\
\sin \theta \sin \psi & \sin \theta \cos \psi & \cos \theta
\end{array}\right)
\end{align}

For the systems studied here---thin films of rod-like molecules deposited at normal incidence---the analysis is greatly simplified by the film's symmetries. The in-plane isotropy ($C_{\infty v}$ symmetry) means the distribution is independent of the in-plane angle $\phi$, which can be treated as uniformly distributed ($f(\phi)=1/2\pi$). To illustrate how this simplifies the orientational average, consider the example of the $\chi^{(2)}_{zxx}$ element for a rod-like molecule:

\begin{align*}
    \chi^{(2)}_{zxx}&=N (l_{z}^{2\omega}l_{x}^{\omega}l_{x}^{\omega} )\beta_{ZZZ}\langle cos(\theta)sin^2(\theta)sin^2(\phi)\rangle\\&=N (l_{z}^{2\omega}l_{x}^{\omega}l_{x}^{\omega} ) \beta_{ZZZ}\int_{0}^{\pi}cos(\theta)sin^2(\theta)f(\theta)d\theta   \int_{0}^{2\pi}\frac{sin^2(\phi)}{2\pi}d\phi\\
    &=N (l_{z}^{2\omega}l_{x}^{\omega}l_{x}^{\omega} ) \beta_{ZZZ}\int_{0}^{\pi}cos(\theta)sin^2(\theta)f(\theta)d\theta \cdot \frac{1}{2}\\&=N (l_{z}^{2\omega}l_{x}^{\omega}l_{x}^{\omega} )\frac{\beta_{ZZZ}}{2}\langle cos(\theta)sin^2(\theta)\rangle\\&=N (l_{z}^{2\omega}l_{x}^{\omega}l_{x}^{\omega} )\frac{\beta_{ZZZ}}{2}(\langle cos(\theta)\rangle-\langle cos^3(\theta)\rangle)
\end{align*}

This same procedure can be applied to all other tensor elements. Because of the rotational symmetry of the rod-like molecules themselves, any dependence on the third Euler angle, $\psi$, also averages out. The final result is that each non-zero element of the $\chi^{(n)}$ tensor can be expressed as a linear combination of the moments of the out-of-plane orientation distribution, $\langle\cos^k\theta\rangle$, up to order $k=n+1$. 

\subsection*{S2. Local Field Factor }

The local field factors\cite{boyd2020nonlinear}, $l_u^\omega$, were calculated using the Lorentz-Lorenz model adapted for a uniaxial medium\cite{Kuzyk_Singer_Stegeman_2013}. For a given optical frequency, $\omega$, the factors depend on the direction within the film. The out-of-plane component ($u=z$) is determined by the extraordinary refractive index, $n_e(\omega)$, while the in-plane components ($u=x,y$) are determined by the ordinary refractive index, $n_o(\omega)$:

\begin{align}
    l_z^\omega &= \frac{\epsilon_e(\omega) + 2}{3} = \frac{(n_e(\omega))^2 + 2}{3} \\
    l_x^\omega = l_y^\omega &= \frac{\epsilon_o(\omega) + 2}{3} = \frac{(n_o(\omega))^2 + 2}{3}
\end{align}

The frequency-dependent extraordinary ($n_e$) and ordinary ($n_o$)  refractive indices used in these calculations were determined from the VASE measurements.

\subsection*{ S3. Explicit Form of the Nonlinear Susceptibility Tensors}

The $\text{n}^{th}$-order nonlinear susceptibility tensor are given here explicitely in terms of moments.  The product of the relevant local field factors $l_{i_0}^{n\omega}\left( \prod_{j=1}^{n} l_{i_j}^\omega \right)$, is  grouped into a single coefficient $L_{i_0i_1..i_n}$.

For the second-order tensor, we define $A=\frac{1}{2}( \langle cos\theta \rangle - \langle cos^3\theta\rangle )$ and $B=   \langle cos^3\theta\rangle$ to obtain: 
\begin{align}
\chi^{(2)} = N \beta_{zzz}
\begin{bmatrix}
0 & 0 & 0 & 0 & L_{xzx} A & 0 \\
0 & 0 & 0 & L_{xzx} A & 0 & 0 \\
L_{zxx} A & L_{zxx} A & L_{zzz} B & 0 & 0 & 0
\end{bmatrix}
\end{align}
The convention for the source field  used to write $\chi^{(2)}$ is: xx, yy,zz,yz,xz,yz.\\

For the third-order tensor, we define $C=\langle \cos^4\theta\rangle$, $D=\frac{3}{8}( 1 - 2\langle \cos^2\theta\rangle + \langle \cos^4\theta\rangle )$, and $E=\frac{1}{2}(\langle \cos^2\theta\rangle - \langle \cos^4\theta\rangle )$ to obtain:
\begin{align}
\chi^{(3)} = N \gamma_{zzzz}
\begin{bmatrix}
L_{xxxx} D & 0& 0 & 0 & 0 & L_{xxzz}E &0 &L_{xxxx}\frac{D}{3} &0 &0  \\
0 & L_{xxxx} D & 0 & L_{xxzz}E  & 0 & 0&0 &0 &L_{xxxx}\frac{D}{3} &0 \\
0 & 0 & L_{zzzz} C & 0 & L_{zxxz}E & 0&L_{zxxz}E &0 &0 &0 \\
\end{bmatrix}
\end{align}
The convention for the source field  used to write $\chi^{(3)}$ is:xxx, yyy, zzz, yzz, yyz, xzz,xxz,xyy,xxy,xyz.\\

For the fourth-order tensor, we define $F = \langle\cos^5\theta\rangle$, $G = \frac{1}{2}(\langle\cos^3\theta\rangle - \langle\cos^5\theta\rangle)$, and $H = \frac{3}{8}(\langle\cos\theta\rangle - 2\langle\cos^3\theta\rangle + \langle\cos^5\theta\rangle)$. 

We can further define:
\begin{align*}
    \chi^{(4)}_{zzzzz}&= N\delta_{zzzzz} L_{zzzzz}F\\
    \chi^{(4)}_{zxxxx}&= N\delta_{zzzzz} L_{zxxxx} H\\
    \chi^{(4)}_{xxxxz}&= N\delta_{zzzzz} L_{xxxxz} H\\
    \chi^{(4)}_{xxzzz}&= N\delta_{zzzzz} L_{xxzzz} G\\
    \chi^{(4)}_{zxxzz}&= N\delta_{zzzzz} L_{zxxzz} G\\
\end{align*}

The full tensor is given in two parts. For columns 1-8:
\[
\chi^{(4)}_{1-8} 
\begin{bmatrix}
    0 & 0 & 0 & 0 & \chi^{(4)}_{xxzzz} & \chi^{(4)}_{xxxxz} & 0 & 0 \\
    0 & 0 & 0 & \chi^{(4)}_{xxxxz} & 0 & 0 & \chi^{(4)}_{xxzzz} & 0 \\
    \chi^{(4)}_{zxxxx} & \chi^{(4)}_{zxxxx} & \chi^{(4)}_{zzzzz} & 0 & 0 & 0 & 0 & 0
\end{bmatrix}
\]
and for columns 9-15:
\[
\chi^{(4)}_{9-15} = 
\begin{bmatrix}
    0 & 0 & 0 & 0 & 0 & \frac{\chi^{(4)}_{xxxxz}}{3} & 0 \\
    0 & 0 & 0 & 0 & \frac{\chi^{(4)}_{xxxxz}}{3} & 0 & 0 \\
    0 & \chi^{(4)}_{zxxzz} & \chi^{(4)}_{zxxzz} & \frac{\chi^{(4)}_{zxxxx}}{3} & 0 & 0 & 0
\end{bmatrix}
\]

The convention for the source field  used to write $\chi^{(4)}$ is: xxxx, yyyy, zzzz, yyyz, xzzz, xxxz, yzzz, xxxy, xyyy, yyzz, xxzz, xxyy, xxyz, xyyz, xyzz. \\

\subsection*{S4. Nonlinear polarimetry data at all angles}

\begin{figure}[h]
    \centering
    \includegraphics[width=0.9\textwidth]{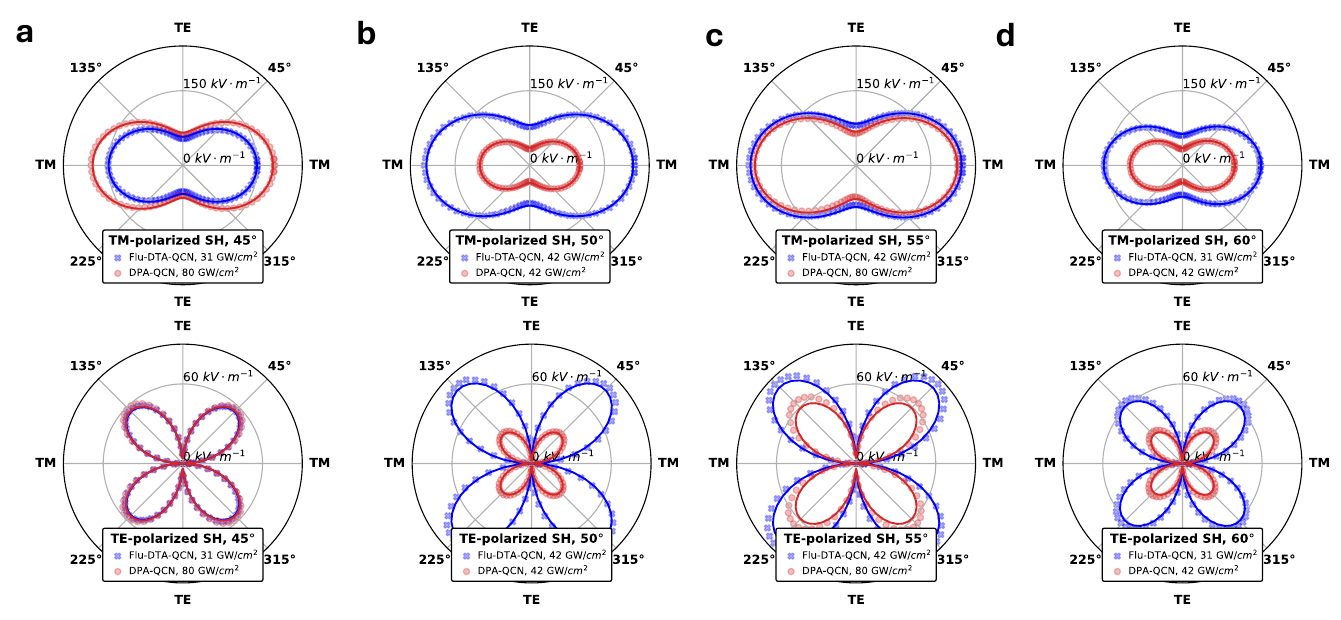} 
\caption{\textbf{Angle-dependent second-harmonic polarimetry.} Experimental (points) and simulated (solid lines) SHG polarimetry patterns for Flu-DTA-QCN and DPA-QCN measured at incidence angles of \textbf{(a)} 45\textdegree, \textbf{(b)} 50\textdegree, \textbf{(c)} 55\textdegree, and \textbf{(d)} 60\textdegree. The top row shows the transverse magnetic (TM) polarized SHG signal, and the bottom row shows the transverse electric (TE) polarized signal. Data is plotted as electric field amplitude. The solid lines represent the global fit obtained using the nonlinear transfer matrix model described in the text. Incident peak intensities are indicated in the legends.}
    \label{fig:SHG_all_angles}
\end{figure}

\begin{figure}[h]
    \centering
    \includegraphics[width=0.9\textwidth]{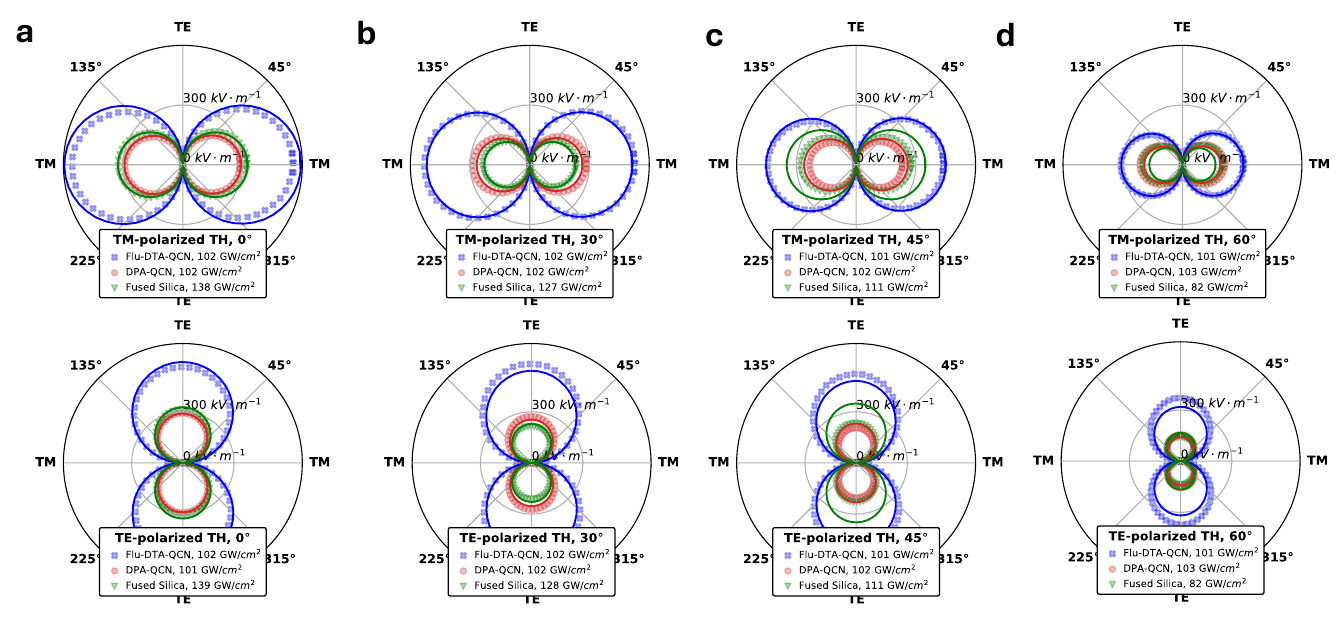} 
\caption{\textbf{Angle-dependent third-harmonic polarimetry.} Experimental (points) and simulated (solid lines) THG polarimetry patterns for Flu-DTA-QCN, DPA-QCN and the fused silica substrate measured at incidence angles of \textbf{(a)} 0\textdegree, \textbf{(b)} 30\textdegree, \textbf{(c)} 45\textdegree, and \textbf{(d)} 60\textdegree. The top row shows the transverse magnetic (TM) polarized SHG signal, and the bottom row shows the transverse electric (TE) polarized signal. Data is plotted as electric field amplitude. The solid lines represent the global fit obtained using the nonlinear transfer matrix model described in the text. Incident peak intensities are indicated in the legends.}
    \label{fig:THG_all_angles}
\end{figure}

\begin{figure}[h]
    \centering
    \includegraphics[width=0.45\textwidth]{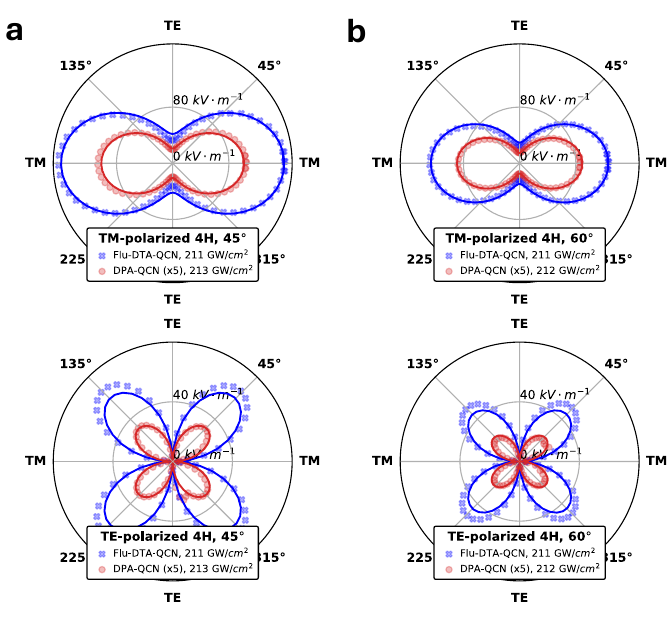} 
\caption{\textbf{Angle-dependent fourth-harmonic polarimetry.} Experimental (points) and simulated (solid lines) 4HG polarimetry patterns for Flu-DTA-QCN, DPA-QCN measured at incidence angles of \textbf{(a)} 45\textdegree, and \textbf{(b)} 60\textdegree. The top row shows the transverse magnetic (TM) polarized SHG signal, and the bottom row shows the transverse electric (TE) polarized signal. Data is plotted as electric field amplitude. For DPA-QCN, the signal is multiplied by 5 to facilitate the comparison. The solid lines represent the global fit obtained using the nonlinear transfer matrix model described in the text. Incident peak intensities are indicated in the legends.}
    \label{fig:4HG_all_angles}
\end{figure}

Figures \ref{fig:SHG_all_angles}, \ref{fig:THG_all_angles}, and \ref{fig:4HG_all_angles} present the complete second-, third-, and fourth-harmonic polarimetry datasets, respectively, measured at multiple incidence angles. The solid lines represent the results of the global fit, demonstrating the model's ability to simultaneously reproduce the nonlinear response across all angular conditions.

\subsection*{S5.  Moments Bounds}

The molecular orientation distribution, $f(\cos\theta)$, is a probability density function of a single variable $x=cos\theta$ defined on the interval $[-1, 1]$. The $n$-th moment of this distribution is defined as:
\begin{equation}
M_n = \langle x^n \rangle = \int_{-1}^{1} x^n f(x) dx
\end{equation}
By definition, the distribution function is non-negative, $f(x) \ge 0$, and normalized, such that the zeroth moment $M_0 = \int_{-1}^{1} f(x) dx = 1$.

The bounds on any moment $M_k$ are not arbitrary but are strictly constrained by the values of the lower-order moments $\{M_0, M_1, \dots, M_{k-1}\}$. To ensure the recovered moments correspond to a valid probability distribution, they must satisfy positivity conditions derived from the \textit{Truncated Hausdorff Moment Problem} \cite{Schmudgen_2017book}. All the mathematical basis for the determination of these bounds can be found in Ref. \citenum{Schmudgen_2017}.

These conditions are checked using three specific types of symmetric matrices, defined directly by their elements $[A]_{i,j}$ (where indices $i,j$ start at 0):
\begin{enumerate}
    \item \textbf{Standard Hankel ($H_k$):} $[H_k]_{i,j} = M_{i+j}$
    \item \textbf{Sum/Diff Shifted ($H^\pm_k$):} $[H^\pm_k]_{i,j} = M_{i+j} \pm M_{i+j+1}$
    \item \textbf{Boundary Shifted ($H^{sq}_k$):} $[H^{sq}_k]_{i,j} = M_{i+j} - M_{i+j+2}$
\end{enumerate}

We determine the bounds recursively. Assuming the moments up to $M_{n-1}$ are already valid, the bounds for the next moment $M_n$ are found by requiring the determinant of the relevant matrices to be non-negative ($\det \ge 0$). This condition is sufficient because the associated Hankel matrices must be positive semidefinite, and (given the validity of lower orders) the non-negativity of the determinant ensures this property is preserved\cite{strang2012linear}.

The specific matrices required depend on whether the moment $n$ is even or odd\cite{Schmudgen_2017}. For an even moment (n = 2k), the lower bound is set by the Standard Hankel matrix $H_{k}$, and the upper bound is set by the Boundary Shifted matrix $H^{sq}_{k-1}$. For an odd moment (n = 2k+1), the Standard Hankel matrix is not used. Instead, the lower bound is set by the Sum Shifted matrix $H^+_k$, and the upper bound is set by the Diff Shifted matrix $H^-_k$.

\subsubsection*{Bounds on $\mathbf{M_1}$ ($\mathbf{k=0}$)}
\begin{itemize}
    \item Lower Bound: $\det(H^+_0) = \det \begin{pmatrix} M_0+M_1  \end{pmatrix} \ge 0$ 
$$ \implies \boxed{M_1 \ge {-M_0}}$$
    \item Upper Bound: $\det(H^{-}_0) = \det \begin{pmatrix} M_0 - M_1\end{pmatrix} \ge 0 $
    $$ \implies \boxed{M_1 \le M_0}$$
\end{itemize}

\subsubsection*{Bounds on $\mathbf{M_2}$ ($\mathbf{k=1}$)}
\begin{itemize}
    \item Lower Bound: $\det(H_1) = \det \begin{pmatrix} M_0 & M_1 \\ M_1 & M_2 \end{pmatrix} \ge 0$ 
    
$$ \implies \boxed{M_2 \ge \frac{M_1^2}{M_0}}$$
    \item Upper Bound: $\det(H^{sq}_0) = \det \begin{pmatrix} M_0 - M_2 \end{pmatrix} \ge 0 $
    $$ \implies \boxed{M_2 \le M_0}$$
\end{itemize}

\subsubsection*{Bounds on $\mathbf{M_3}$ ($\mathbf{k=1}$)}
\begin{itemize}
    \item Lower Bound: $\det(H^+_1) = \det \begin{pmatrix} M_0+M_1 & M_1+M_2 \\ M_1+M_2 & M_2+M_3 \end{pmatrix} \ge 0$.
    Solving for $M_3$:
    \begin{equation*}
    \boxed{M_3 \ge \frac{(M_1 + M_2)^2}{M_0 + M_1} - M_2}
    \end{equation*}
    \item Upper Bound: $\det(H^-_1) = \det \begin{pmatrix} M_0-M_1 & M_1-M_2 \\ M_1-M_2 & M_2-M_3 \end{pmatrix} \ge 0$.
    Solving for $M_3$:
    \begin{equation*}
    \boxed{M_3 \le M_2 - \frac{(M_1 - M_2)^2}{M_0 - M_1}}
    \end{equation*}
\end{itemize}

\subsubsection*{Bounds on $\mathbf{M_4}$ ($\mathbf{k=2}$)}
\begin{itemize}
    \item Lower Bound: $\det(H_2) \ge 0$.
    \begin{equation*}
    \det \begin{pmatrix} M_0 & M_1 & M_2 \\ M_1 & M_2 & M_3 \\ M_2 & M_3 & M_4 \end{pmatrix} \ge 0
    \end{equation*}
    $$ \implies \boxed{M_4 \ge \frac{M_0M_3^2 +M_2^3-2M_1M_2M_3}{M_0M_2-M_1^2}}$$
    \item Upper Bound: $\det(H^{sq}_1) \ge 0$.
    \begin{equation*}
    \det \begin{pmatrix} M_0 - M_2 & M_1 - M_3 \\ M_1 - M_3 & M_2 - M_4 \end{pmatrix} \ge 0 \end{equation*} $$\implies \boxed{M_4 \le M_2 - \frac{(M_1 - M_3)^2}{M_0 - M_2}}$$
\end{itemize}

\subsubsection*{Bounds on $\mathbf{M_5}$ ($\mathbf{k=2}$)}
\begin{itemize}
    \item Lower Bound: $\det(H^+_2) \ge 0$.
    \begin{equation*}
    \det \begin{pmatrix} 
    M_0+M_1 & M_1+M_2 & M_2+M_3 \\ 
    M_1+M_2 & M_2+M_3 & M_3+M_4 \\ 
    M_2+M_3 & M_3+M_4 & M_4+M_5 
    \end{pmatrix} \ge 0
    \end{equation*}
    \item Upper Bound: $\det(H^-_2) \ge 0$.
    \begin{equation*}
    \det \begin{pmatrix} 
    M_0-M_1 & M_1-M_2 & M_2-M_3 \\ 
    M_1-M_2 & M_2-M_3 & M_3-M_4 \\ 
    M_2-M_3 & M_3-M_4 & M_4-M_5 
    \end{pmatrix} \ge 0
    \end{equation*}
\end{itemize}
Closed-form expressions for these bounds are available but are omitted here for brevity; 
they can be straightforwardly obtained by expanding the determinant of the $3 \times 3$ matrices 
and solving the resulting linear inequality for $M_5$.

\subsection*{S6. Transformation of Orientation Distributions}

To analyze the anisotropy of the transition dipole moments, the molecular orientation distributions initially calculated as a function of $\cos\theta$ are transformed into probability density functions of $u = \cos^2\theta$.

The probability density function $f(u)$ is derived by summing the contributions from both the positive ($+\sqrt{u}$) and negative ($-\sqrt{u}$) branches of the original distribution. By the conservation of probability $f(u)du = f(x)dx$, the transformation is given by:

\begin{equation}
    f(u) = \left| \frac{dx}{du} \right| \left[ f(\sqrt{u}) + f(-\sqrt{u}) \right]
\end{equation}

Applying the Jacobian of the coordinate transformation $|dx/du| = (2\sqrt{u})^{-1}$, the final density function becomes:

\begin{equation}
    f(u) = \frac{f(\sqrt{u}) + f(-\sqrt{u})}{2\sqrt{u}}
    \label{eq:transform}
\end{equation}

where $u = \cos^2\theta$. The singularity observed as $\cos^2\theta \to 0$ is a mathematical consequence of the Jacobian term (the vanishing gradient of the quadratic function at zero).

The uncertainty (standard deviation), denoted as $\sigma$, was propagated assuming the errors at $+\sqrt{u}$ and $-\sqrt{u}$ are uncorrelated. Following standard error propagation rules for the sum in Eq. \ref{eq:transform} scaled by the Jacobian factor, the transformed uncertainty is calculated as:

\begin{equation}
    \sigma_{f(u)} = \frac{\sqrt{\sigma^2(\sqrt{u}) + \sigma^2(-\sqrt{u})}}{2\sqrt{u}}
\end{equation}

\begin{figure}[h]
    \centering
    \includegraphics[width=0.6\textwidth]{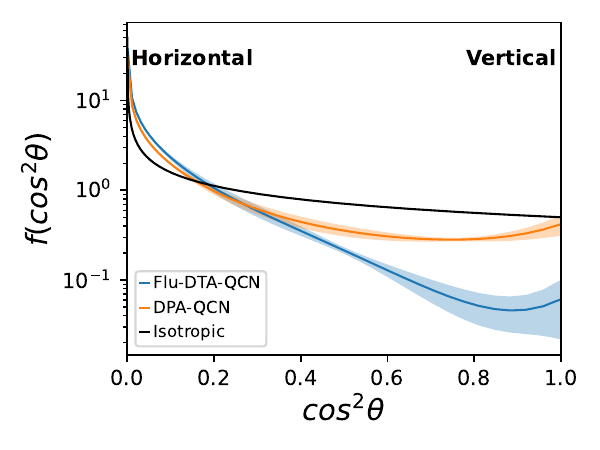} 
\caption{\textbf{Transformed molecular orientation distributions.} The probability density functions are plotted as a function of $\cos^2\theta$ on a logarithmic scale. Values near $\cos^2\theta \approx 0$ indicate alignment within the surface plane (Horizontal), while values near $\cos^2\theta \approx 1$ indicate alignment along the surface normal (Vertical). The shaded regions represent the standard deviation propagated from the original distributions. For reference, a completely isotropic distribution (random orientation) would follow the analytical curve $f_{\text{iso}}(u) = (2\sqrt{u})^{-1}$, yielding an expectation value of $\langle \cos^2\theta \rangle = 1/3$.}
    \label{fig:cos2_dist}
\end{figure}

Figure \ref{fig:cos2_dist} presents the transformed orientation distributions for Flu-DTA-QCN and DPA-QCN. The solid black line represents the theoretical isotropic distribution ($f_{\text{iso}}(u) = 1/(2\sqrt{u}$)), which yields an expectation value of $\langle \cos^2\theta \rangle = \int_0^1 u f_{\text{iso}}(u) du = 1/3$. Deviations from this baseline indicate preferential alignment. 

Both species exhibit a probability density exceeding the isotropic prediction near $\cos^2\theta \approx 0$, as anticipated from the preference for horizontal alignment. In the case of DPA-QCN, the slight recovery in density as $\cos^2\theta \to 1$ implies a weak bimodal character; however, the horizontal component remains significantly stronger than the vertical component.

\subsection*{S7. Molecular dynamics simulations: Details}
\textbf{Customized Force-field parameters}
The non-bonded interactions between the atoms of the deposited molecules and the substrate were slightly modified ($\epsilon$ was doubled for the atom-types CA, CM, NC and HA). This small modification resulted in improved agreement with experimental results. 

\textbf{Glass transition temperature $T_g$ in simulations}
The glass transition temperatures ($T_g$) for the two materials (DPA-QCN and Flu-DTA-QCN), were identified from the inflection point in temperature dependence of molecular diffusion. The diffusion coefficients were calculated using the Green–Kubo approach. To this end, we ran MD simulations to deposit 500 molecules on the graphene substrate following the deposition protocol described in the main text in a broad temperature range (283 K to 562 K) and followed up with a brief equilibration (2 ns). The center-of-geometry velocities for 300 molecules selected at random were recorded in the last 3 ps of equilibration runs at every time-step (1 fs) and used to construct the velocity autocorrelation function (VACF). The diffusion coefficients at each temperature were then calculated by numerically integrating the VACF. The low/high temperature regimes were identified by choosing consecutive points for each group while optimizing the combined quality of the bilinear fits. The temperature at the intersection of the linear regimes was interpreted as the glass transition temperature . For Flu-DTA-QCN $T_g = 439$ K and for DPA-QCN $T_g = 549.3$ K. These values describe the transition observed in simulations and they are affected by the accuracy of the force-field and other simulation parameters; as such, they are not expected or required to be very close to the experimental values. In order to approximate experimental conditions, the deposition simulations analyzed in this work were run at temperatures set to 0.8$T_g$.

\subsection*{S8. Moments of the simulated distribution}

Table S\ref{tab:sim_moments} show the moments extracted from the MD-simulated distributions. The moments are calculated the binned (bin size of 0.02) data.

\begin{table}[ht]
    \centering
    \begin{tabular}{ccc}\hline
       \textbf{Moment}  & \textbf{Flu-DTA-QCN} & \textbf{DPA-QCN} \\ \hline 
    $\langle cos \theta \rangle $ &0.022 &0.0484 \\ 
    $\langle \cos^2 \theta \rangle $ &0.19 & 0.265\\
    $\langle \cos^3 \theta \rangle$ & 0.0066 &0.0257 \\ 
     $\langle \cos^4 \theta \rangle$ & 0.099&0.143\\
      $\langle \cos^5 \theta \rangle$ & 0.0052&0.0172\\\hline
    \end{tabular}
    \caption{The first five moments of the MD-simulated molecular orientation distribution for Flu-DTA-QCN and DPA-QCN.}
\label{tab:sim_moments}

\end{table}